\documentclass[preprint2]{aastex}
\usepackage{graphicx}
\usepackage{epstopdf}
\usepackage{footnote}

\shorttitle{\textit{INTEGRAL}/SPI Timing Analysis of V404 Cyg}
\shortauthors{Rodi et al.}

\begin{document}


\title{Timing Analysis of V404 Cyg during Its Brightest \\ Outburst with \textit{INTEGRAL}/SPI}


\author{J. Rodi\altaffilmark{1,2}, E. Jourdain\altaffilmark{1,2}, and J. P. Roques\altaffilmark{1,2}}
\affil{\textsuperscript{1}Universit\'e de Toulouse; UPS-OMP; IRAP;  Toulouse, France\\
\textsuperscript{2}CNRS; IRAP; 9 Av. Colonel Roche, BP 44346, F-31028 Toulouse cedex 4, France\\}

 
\begin{abstract}
The outburst of V404 Cyg during the summer of 2015 reached unparalleled intensities at X-ray and soft gamma-ray energies with fluxes \( > 50\) Crab in the \(20-50\) keV energy band.  To date, studies in the hard X-ray/soft gamma-ray energy domain have focused primarily on the energy spectra.  In this work, timing analysis has been performed with \textit{INTEGRAL}/SPI data in the \(20-300\) keV energy range for \textit{INTEGRAL} revolution 1557, which corresponds to the brightest flare of V404 Cyg (on June, 26). The power spectra are fit with broken power-law and multi-Lorentzian models and compared with previously reported results of V404 Cyg flaring activity from 1989 and 2015.  Also, we took advantage of the good signal-to-noise ratio obtained above 70 keV to quantify the timing/fast-variability properties of the source as a function of energy.  We then point out similarities of V404 Cyg with the black hole transient V4641 Sgr. Like V4641 Sgr, we found the power spectra of V404 Cyg during high flux periods did not possess the expected flat-top feature typically seen in a hard spectral state.  Interpretations are proposed in the framework of the fluctuating-propagation model to explain the observed properties.

\end{abstract}




\section{Introduction}

On 2015 June 15 at 18:31:38 UTC (MJD 57188.772), the \textit{Swift}/Burst Alert Telescope (BAT) triggered on the black hole (BH) transient V404 Cyg (GS \(2023+338\)) \citep{barthelmy2015}, which began a \( \sim 1\) month-long period of strong activity with unprecedented coverage at all wavelengths.  In particular in the hard X-rays, intensities reached values \(>50\) Crab in the \(20-50\) keV band \citep{jourdain2016}.  So far, most analyses at hard X-ray/soft gamma-ray energies have focused primarily on the energy spectra, which have found the source spectrum well-described by a Comptonization model, plus a second component to take into account the high energy tail \citep{roques2015,natalucci2015,jenke2016,jourdain2016}.  Together with an absent or moderate thermal (disk) emission (see for instance \cite{radhika2016}), this suggests the source to be mainly in Hard (or Intermediate Hard) State.  However, in these works, Comptonization models have found best-fit parameters difficult to explain, reinforcing the atypical behavior of  V404 Cyg mentioned since its 1989 outburst  \citep{oosterbroek1997}.  The reported seed photon temperatures are \(\sim 5-7\) keV \citep{roques2015,natalucci2015,jenke2016,jourdain2016}, which are significantly hotter than the \( \sim 1\) keV that can be produced by an accretion disk.  It could suggest that the seed photons are due to another process such as bremsstrahlung or synchrotron emission \citep{markoff2005,done2007}.  On the other hand, absorption values of \( \sim 1-\) a few \(10^{24} \textrm{ cm}^{-2}\) \citep{jenke2016,jourdain2016, motta2017} and/or a very strong reflection component may provide a good description of the data, confirming the complex variable geometry reported during the previous outburst \citep{oosterbroek1997,zycki1999}.

Other properties of the systems, such as accretion geometry and the dynamics of the Comptonizing medium, can be studied through timing analysis \citep{kazanas1997,reig2003,giannios2004}.  While many instructive results are available in the X-ray domain, results in the hard X-ray regime are few, where photons are much less numerous.  Such results are particularly important to investigate the emission mechanism at work above 20 keV.

In this work, we report on timing analysis of \textit{INTEGRAL}/SPI data in the \(20-300\) keV energy range, around the brightest period of the V404 Cyg outburst.  First, we describe the observations along with the analysis method.  Next, we present our power spectra as well as time lags between low and high energy bands.  Finally, we discuss our results in context of other results from V404 Cyg and the BH transient V4641 Sgr.

\begin{table*}
\begin{center}
\caption{\textit{INTEGRAL}/SPI observations of V404 Cyg during Revolution 1557 used in this work}
\begin{tabular}{cccc}
\tableline\tableline \\
Period &                              & Time                         &  Time   \\
       &                              & (MJD)                        &  (UTC)  \\  
\tableline
1      &  Flare 1                     & \(57199.110-57199.250\)      & 2015 June 26 02:38:24 \(-\) 06:00:00  \\
2      &  Flare 2 Rise                & \(57199.470-57199.510\)      & 2015 June 26 11:16:48 \(-\) 12:14:24  \\
3      &  Flare 2 Decay               & \(57199.845-57199.860\)      & 2015 June 26 20:16:48 \(-\) 20:38:24  \\
4      &  Flare 2 Tail                & \(57200.082-57200.148\)      & 2015 June 27 01:58:04 \(-\) 03:33:07  \\
\tableline\tableline
\label{table:obs}
\end{tabular}

\label{tab:obs}
\end{center}
\end{table*}

\begin{figure*}[t!]
  \centering
  \includegraphics[scale=0.65, angle=180,trim = 0mm 20mm 23mm 0mm, clip]{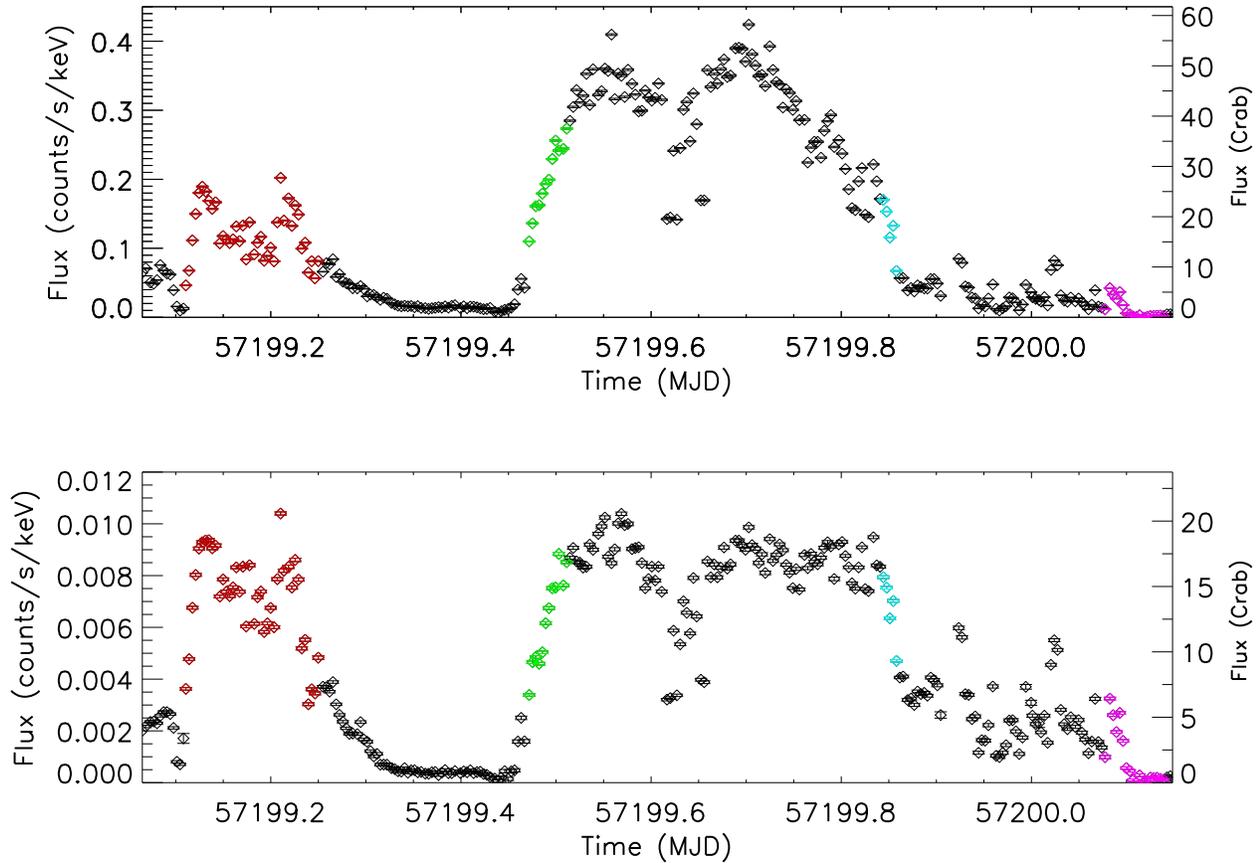}
  \caption{\textit{Top}: \textit{INTEGRAL}/SPI light curve of V404 Cyg in \(20-50\) keV band spanning MJD \(57199.064-57200.150\).  Points are 300 s integration time.  The red, green, and blue periods are high SNR times that do not suffer from telemetry loss.  The purple period allows for the study of a low flux period and is also contemporaneous with \textit{Fermi}/GBM observations.  \textit{Bottom}: Light curve of V404 Cyg in \(50-300\) keV energy band with the observation periods denoted with the same colors as above.}  \label{fig:lc}
\end{figure*}

\section{Observations and Data Analysis}
\textit{INTEGRAL}/SPI \citep{vedrenne2003,roques2003} observations of V404 Cyg during the outburst maximum occurred during revolution 1557, which spanned  2015 June 26 01:32 \(-\) 27 23:47 UTC (MJD \(57199.064-57200.991\)).  During this revolution, V404 Cyg underwent two major flares, as shown in Fig.~\ref{fig:lc}.  The light curves cover the \(20-50\) keV (top) and \(50-300\) keV (bottom) energy bands with 300 s integration for each point. 

A study of the temporal behavior of V404 Cyg was conducted using four time periods during which the source was extremely bright to maximize the signal-to-noise ratio (SNR) and to minimize any contamination from Cyg X-1, which is in the SPI field of view.  However, the loss of telemetry packets during the brightest times of the second flare, even though moderate (\( \sim 10-15 \%\)), made timing analysis difficult thus data from this time span were not analyzed. The data sets studied in this work are plotted in red, green, blue, and purple, respectively, with the start and stop times listed in Table~\ref{table:obs}.  The first period (red) spans the majority of the first flare, beginning when the \(20-50\) keV flux exceeds \(\sim 5\) Crab after a short-lived drop and lasting until the flare begins its smooth decay to a ``low'' flux state.  The second period (green) covers the rise of the second flare while the third period (blue) consists of the decay of the second flare.  The last period (purple) corresponds to the tail of the flare and is contemporaneous with \textit{Fermi}/GBM observations used by \citet{jenke2016}. It marks the end of the period of exceptional activity.

For each period, power density spectra (PDS) were made in three energy ranges.  A broad \(20-300\) keV energy range was studied first to investigate the general shape of the power spectra.  Then two narrower energy bands, \(20-50\) keV and \(69-300\) keV, were analyzed to study the behavior as a function of energy.  Because of intense background lines \citep{weidenspointner2003}, the \(50-69\) keV and \(190-215\) keV energy ranges were excluded in constructing the high energy band.  The data from individual germanium detectors were summed together for getting the total number of counts incident on the detector plane.  In constructing a single PDS, 100-s long segments were used, and the SPI data, which have a time resolution of \(102.4 \textrm{ } \mu\)s \citep{vedrenne2003}. The data were rebinned to have a Nyquist frequency of 81 Hz (\( f_{Nyq} = 1/(2 \Delta t)\), where \(\Delta t\) is the time bin width; \(\Delta t\) = \(614.4 \textrm{ } \mu s\) here), which allowed for studying frequencies from \(0.01-81\) Hz.  Power spectra analysis was performed using the Interactive Spectral Interpretation System (ISIS) \citep{houck2000}.

\begin{figure}[b!]
  \centering
  \includegraphics[scale=0.75, angle=0,trim = 20mm 70mm 90mm 130mm, clip]{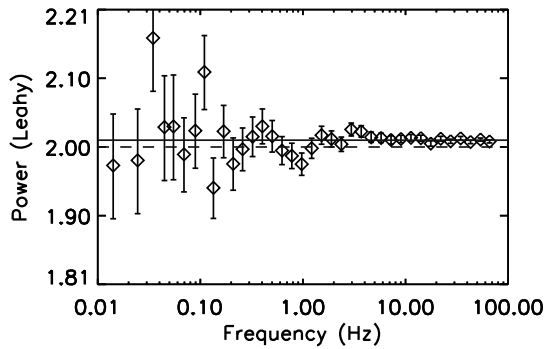}
  \caption{PDS of \(\sim \)70 000 s of empty field observations from revolution 1549 for the \(20-300\) keV energy range.  Solid line plotted at average power at 2.0099;  Dashed line at power of 2.} \label{fig:background}
\end{figure}

PDS of empty field or background observations will still result in a nonzero average value because of statistical variability.  Using the Leahy normalization \citep{leahy1983}, the mean value of the PDS should be 2 in the case of only a background signal.  To verify that the SPI data behave  as expected, data from \( \sim \) 70 000 s of observations from revolution 1549 were used as an empty field observation (centered on \( \alpha = 12:30:00, \textrm{ } \delta= +10:00:00\)).  Fig.~\ref{fig:background} shows the Leahy normalized PDS from the empty field period in the \(20-300\) keV energy range.  The PDS displays a fairly constant power with small scatter from \( \sim 1-81\) Hz.  Below \(\sim 1\) Hz, the scatter increases, though the sizes of the error bars increase as well.  A fit to a constant value finds an average power of the PDS of \(2.0099 \pm 0.0009\) with \( \chi^2 / \nu = 0.889 \) (\( \nu =35\)), which is close to the expected value.  As the SPI data show values close to 2 for the total energy range as well as for the two narrower energy bands, the theoretical value of 2 was used to remove the background component in converting the PDS to rms normalization \citep{belloni1990b,miyamoto1991}.

\section{Results}

\subsection{Power Spectra Density Fits}
Power spectra can often be described by multiple components that are generally classified as either band-limited noise (BLN), which are broad features, or quasi-periodic oscillations (QPOs), which are narrow features \citep{belloni2002}.

Initially, the V404 Cyg data were fit with a two Lorentzian model, but large positive features in the residuals were often found, suggesting the existence of a third component except for observation period 4 in which the SNR is low.  The presence of three broad features is consistent with \citet{belloni2002}, who found similar results in nine of the 10 compact systems they studied, as well as with the analysis from the 1989 outburst of V404 Cyg by \citet{oosterbroek1997}. All the Lorentzians are zero-centered and no QPOs were detected in this work.

\begin{figure*}[t!]
  \centering
\includegraphics[scale=0.95, angle=0,trim = 15mm 0mm 30mm 80mm, clip]{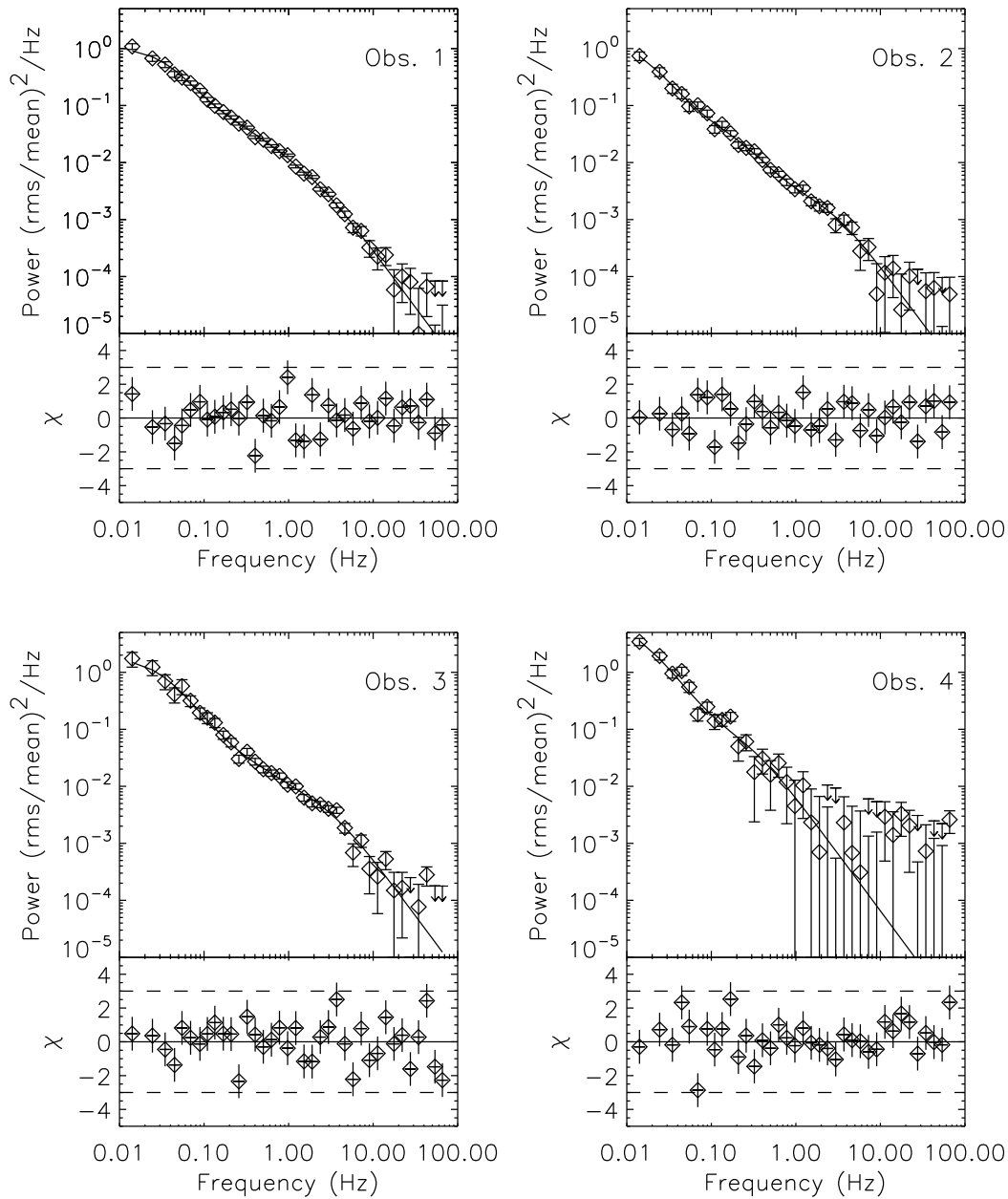}
  \caption{The \(20-300\) keV power spectra of the four observation periods with two or three Lorentzian fit models overplotted.  The corresponding residuals are plotted below.  The solid line denotes \( 0 \sigma\), and the dashed lines denote \( \pm 3 \sigma\).  See Table 2 for the model best-fit parameters.}  \label{fig:l3pdf}
\end{figure*}

Adopting the notation from \citet{belloni2002}, the lowest frequency BLN component, \(L_b\), corresponds to the ``break'' in the power spectrum at frequency \( \nu_b\).  The other two BLN components are fit by the Lorentzians \(L_l\) and \(L_u\) with frequencies \(\nu_l\) and \(\nu_u\) to cover the ``lower'' and ``upper'' parts of the power spectrum.  

The power spectra for the four observation periods are plotted in Fig.~\ref{fig:l3pdf} with the best-fit model overplotted and the residuals plotted below. For the first three observation periods, the power spectra are significantly detected up to \(\sim 5-10\) Hz.  The flux level of V404 Cyg during observation period 4 is drastically lower than for the others so a lower SNR is expected.  The source signal is significantly detected out to only \(\sim 0.5\) Hz while presenting the highest power at low frequencies.

The results of the fits to each power spectrum are listed in Table~\ref{table:l3}, which includes in the top panel the \( \nu_{max}\) and rms for the three (or two) Lorentzians along with the total rms and \( \chi^2 / \nu\) values.  The errors listed are the 1-\(\sigma\) values. 

The peak frequencies of the SPI power spectra have a narrow range for each of the Lorentzians: \( \sim 0.012 - 0.034\) Hz, \(\sim 0.2-0.6\) Hz, and \( \sim 1.8-2.8\) Hz, for \( \nu_b\), \( \nu_l\), and \( \nu_u\), respectively. The break Lorentzian always dominates the total rms, spanning \( \sim 11.5-26\)\%, while \( \textrm{rms}_l\) and \( \textrm{rms}_u\) remain between \( \sim 6 - 11 \)\%. 

\begin{figure}[t!]
  \centering
  \includegraphics[scale=0.8, angle=0,trim = 2mm 10mm 80mm 130mm, clip]{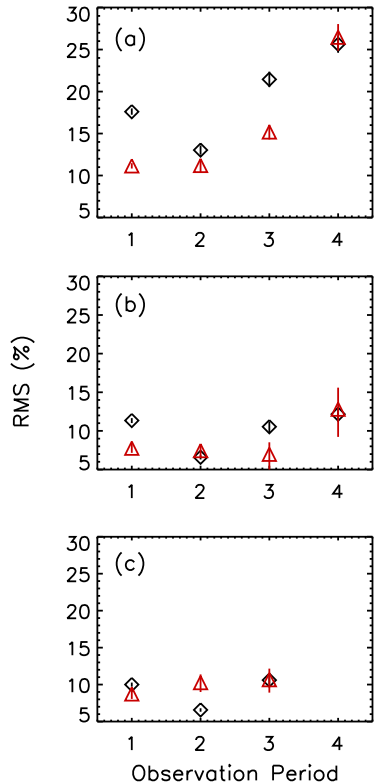}
  \caption{RMS frequency values for L\(_b \), L\(_l \), and L\(_u\) in panels (a), (b), and (c), respectively.  The SPI \(20-50\) keV results are plotted as black diamonds, and the \(69-300\) keV results are plotted in red triangles.}  \label{fig:l3fits}
\end{figure}

Considering the energy domain accessible in our data, we were able to investigate the evolution of the rms values in the hard X-ray domain.  We thus built power spectra for each of the four time periods in two energy bands: \(20-50\) keV and \(69-300\) keV. We fixed the \( \nu_{max}\) frequencies to the \(20-300\) keV ones (Table~\ref{table:l3}) and determined the rms values.  Results are displayed in Fig.~\ref{fig:l3fits}, where panels (a), (b), and (c) correspond to L\(_b\), L\(_l\), and L\(_u\), respectively, while black diamonds and red triangles stand for \(20-50\) keV and \(69-300\) keV bands. For each Lorentzian, the rms values follow a similar pattern, with an upward trend between periods \(2-4\) (i.e. during the second flare). However, the evolution with  energy is different for each period: in periods 1 and 3, the low energy values are larger by a factor of \( \sim 1.5\) for \( \textrm{rms}_b\) and \( \textrm{rms}_l\).  For \( \textrm{rms}_u\), the ratio decreases to \( \sim 1.1\) for period 1 while for period 3, the ratio is approximately 1.  For period 2, the ratio decreases with increasing frequency from \( \sim 1.2\) to \( \sim 0.9\) to \( \sim 0.6\).  For period 4, the rates are close to 1 with consistent values between the low and high energy rms values for both Lorentzians.

\label{bkn}
The power spectra were also fit using broken power-law models for comparison with other results or model predictions. (See Sec.~\ref{1989} and Sec.~\ref{othersources}.)  The fit parameters are listed in the bottom panel Table~\ref{table:l3} with observation 4 requiring only a single power law.  For observations 1, 2 and 4, the \( \chi^2 / \nu\) values are similar to those obtained with multi-Lorentzian models.  Observation 3 residuals suggested the need for a low frequency (\(< \sim 0.1 \) Hz) component.  Adding a third power-law results in a slope of 1.48 below 0.26 Hz while the middle power law hardens to \( \sim 1\), the high-frequency power-law softens to roughly 2.1 and the \( \chi^2 / \nu\) value decreases to 1.43.  If a Lorentzian is included instead, a peak frequency of \(0.023^{+0.008}_{-0.006}\) Hz is found, which is consistent with \(L_b\) in the three Lorentzian model. The final \( \chi^2 / \nu\) value reduces to 1.46.  We note that the inclusion of a third power-law or a Lorentzian results in a better \( \chi^2 / \nu\) value than the multi-Lorentzian model.

\begin{table*}
\begin{center}
\caption{Power Spectra Fits}
\scalebox{1.}{ 
\label{table:l3}
\begin{tabular}{lccccccccccccccc}
\tableline\tableline \\

Obs. period  &                     1&                      &2&                    &3&                   &4&                     \\  
\tableline \\
\multicolumn{8}{c}{Multi-Lorentzian}\\
\hline
\( \nu_b\) (Hz)             &\(0.034^{+0.005}_{-0.007}\)&&\(0.012^{+0.005}_{-0.004}\)&&\(0.029^{+0.008}_{-0.006}\)&&\(0.012^{+0.003}_{-0.003}\) \\
\( \textrm{rms}_b\) (\%)    &\(15.0^  {+0.4  }_{-0.6  }\)&&\(11.5^ {+1.7  }_{-1.0  }\)&&\(18.5 ^{+1.4  }_{-1.3 }\)&&\(25.9 ^{+2.8  }_{-1.9  }\) \\
\( \nu_l\) (Hz)             &\(0.4^   {+0.2  }_{-0.2  }\)&&\(0.21^ {+0.06 }_{-0.05 }\)&&\(0.6  ^{+0.3  }_{-0.3 }\)&&\(0.3  ^{+0.2  }_{-0.1  }\) \\
\( \textrm{rms}_l\) (\%)    &\(9.5^   {+1.3  }_{-1.3  }\)&&\(6.5 ^ {+0.4  }_{-0.4  }\)&&\(7.7  ^{+1.4  }_{-1.7 }\)&&\(10.9 ^{+1.2  }_{-1.4  }\) \\
\( \nu_u\) (Hz)             &\(1.8^   {+0.9  }_{-0.5  }\)&&\(2.4 ^ {+0.7  }_{-0.5  }\)&&\(2.8  ^{+0.8  }_{-0.5 }\)&&\(-\)                   \\
\( \textrm{rms}_u\) (\%)    &\(8.8^   {+1.7  }_{-2.1  }\)&&\(5.9 ^ {+0.3  }_{-0.4  }\)&&\(10.2 ^{+1.0  }_{-1.5 }\)&&\(-\)                   \\
\( \textrm{rms}_{tot}\) (\%) &\(19.8^  {+1.7  }_{-1.9  }\)&&\(14.5^ {+1.7  }_{-1.2  }\)&&\(22.6 ^{+2.1  }_{-2.3 }\)&&\(28.1^{+3.0    }_{-2.3 }\) \\
\( \chi^2 / \nu \)          & 1.04                   && 0.94                  && 1.60                 && 1.27                   \\ 

\tableline \\
\multicolumn{8}{c}{Power Law}\\
\hline
\( \Gamma_1\)              &\(-1.11^{+0.01}_{-0.01}\)&&\(-1.20^{+0.03}_{-0.02}\)&&\(-1.09^{+0.03}_{-0.04}\)&&\(-1.50^{+0.07}_{-0.06}\) \\
\( \nu_B\) (Hz)            &\( 1.9 ^{+0.2 }_{-0.2 }\)&&\( 4.7 ^{+2.4 }_{-1.4 }\)&&\( 3.7 ^{+0.2 }_{-0.2 }\)&&\(-\)                  \\
\( \Gamma_2\)              &\(-1.71^{+0.07}_{-0.07}\)&&\(-1.78^{+0.4 }_{-0.8 }\)&&\(-1.8 ^{+0.2 }_{-0.3 }\)&&\(-\)                  \\
\( \chi^2 / \nu \)         & 1.12                 && 0.99                 && 2.07\tablenotemark{a}                && 1.30   \\ 
\tableline\tableline
\tablenotetext{a}{A better model is proposed in Sec.~3.1.}
\end{tabular}
}

\end{center}
\end{table*}
\subsection{Time Lags}
\begin{figure}[t!]
  \centering 
  \includegraphics[scale=0.75, angle=0,trim = 20mm 0mm 40mm 170mm, clip]{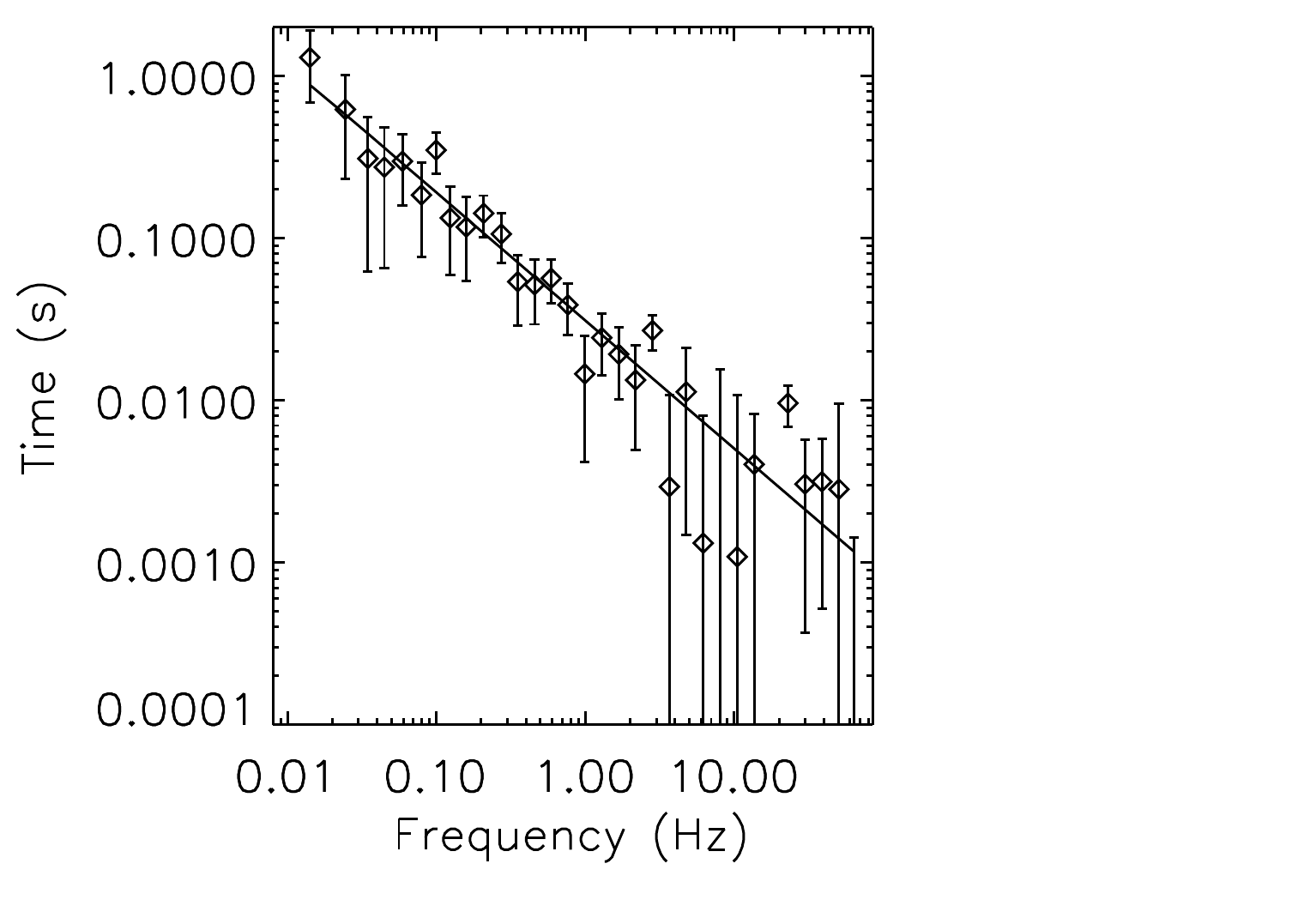}
  \caption{Time lags between the \(20-50\) keV and \(69-300\) keV energy bands from observation period 1.} \label{fig:timelag}
\end{figure}

In addition to generating power spectra, the time lags between the \(20-50\) keV and \(69-300 \) keV energy bands were determined.  Only the first period possessed high enough statistics to make any conclusive statements.  Fig.~\ref{fig:timelag} shows the time lag with positive lags indicating that the hard photons lag the soft ones. The time lags follow a power law over the frequency range, spanning \( \sim 10^{-3} - 1\) s, though above \( \sim 3\) Hz, the lag values are poorly constrained.  The best-fit power law to the data has an index of \( \Gamma = 0.79^{+0.07}_{-0.07} \) (\( \chi^2/ \nu = 1.016\)).

\section{Discussion}
\subsection{SPI results}

Broken power-law and multi-Lorentzian models provide a similarly good description of the data in terms of $\chi^2/ \nu$ values. In both cases, the obtained parameters point out some evolution with time.\\
Periods 1 and 3 are globally alike.  Lorentzian and broken power-law parameters are close, except that an excess of power at low frequencies requires one additional power-law for period 3.  For both periods, \textrm{rms} values are significantly higher at low energy and low frequency. To relate timing properties and spectral ones, we note that period 3 corresponds to the end of the decay of the second flare, during which the emission is observed to recover the same hardness as period 1, through a decrease of the low energy emission (\(20-50\) keV; see  Fig.~\ref{fig:lc} or \citet{jourdain2016}). 

On the other hand, during period 2, (along the rise phase of the second flare),  the spectral evolution is driven by the increase of the low energy part (see Fig.~\ref{fig:lc}). In the time domain, the variability is lower, particularly at low frequency (low \(\textrm{rms}_{tot}\) and \(\textrm{rms}_b\) or steeper \( \Gamma_1\)). Fig.~\ref{fig:l3fits} shows that this decrease is mainly due to low energy: while the \textrm{rms} values are comparable to those of period 1 for the \(69-300\) keV energy band, they are notably smaller in the \(20-50\) keV band. 

Finally, the 4th period behaves quite differently from the others. The lower SNR could explain the absence of signal above 1 Hz in the PDS and a simple power-law describing the PDS spectrum. However, its (low frequency) slope of 1.5 is similar to that of the component added for describing the low frequency part of period 3. It can also be seen as an intermediate value between the low and high frequency indices found in periods 1 and 2.  In addition, because of the comparatively large \( \textrm{rms}_b\), the \(\textrm{rms}_{tot}\) for period 4 is the highest (28.1\%) with respect to the other periods (\(14.5-22.6\)\%). Moreover, as already mentioned, the \( \textrm{rms}\) values are unchanged from 20 to 300 keV.  It should be noted that this period corresponds to a spectrally more stable state, where the Compton emission is decreasing while the hard tail component takes over, producing a particularly hard spectral shape \citep{jourdain2016}.

\subsection{Comparison with Other Observations}
Due to the unprecedented intensities at X/hard X-ray energies and the duration of the outburst, other Fourier-based studies have been performed on V404 Cyg data.  Observations from the \textit{Swift}/X-Ray Telescope (XRT) during MJD 57191 show a weak QPO detection at 1.7 Hz in the \(0.3-10\) keV band during a low flux period \citep{motta2015a}.  \citet{huppenkothen2016} have reported detecting several QPOs at frequencies of 18 mHz, 73 mHz, 136 mHz, and 1.03 Hz with \textit{Fermi}/GBM, \textit{Swift}/XRT, and the \textit{Chandra}/Advanced CCD Imaging Spectrometer (ACIS).  Conversely, \citet{jenke2016} analyzed the \textit{Fermi}/Gamma-ray Burst Monitor (GBM) data from MJD \(57188-57200\) and were able to fit the data with two or three broad Lorentzians and do not report any QPOs.  Also, \citet{radhika2016} reported no detection of QPOs during 2015 June observations with \textit{Swift}/XRT.  Similar to the \citet{jenke2016} GBM results, we were able to describe the SPI PDS with two or three broad Lorentzians with no QPOs detected.

Concurrent observations between \textit{INTEGRAL} and GBM allowed for a comparison between the two instruments.  GBM analysis reported a contemporaneous observation with period 4 (MJD \(57200.082-57200.148\)).  \citet{jenke2016} used \(8-100\) keV data, extending to lower energies than the SPI data.  Data sets from both instruments required only two BLN components.  For the \( \nu_b\) values, \citet{jenke2016} found a best-fit value of \((1.6 \pm 0.5) \times 10^{-2}\) Hz compared to \((1.2 \pm 0.3) \times 10^{-2}\) Hz in the SPI \(20-300\) keV results and thus the results are marginally consistent.  The \( \nu_l\) values are significantly different with \citet{jenke2016} finding \(0.61 \pm 0.1\) Hz while this analysis found \(0.3 \pm 0.1 \) Hz.  A fit to the SPI \(20-300\) keV power spectrum with the frequencies fixed to the GBM values finds a \( \chi^2 / \nu \) of 1.29, only slightly higher than the 1.27 found when both frequencies are left free, but the \( \chi^2\) is larger by 3.01.  However, the small disagreement is potentially  due to the different energy ranges analyzed from the two instruments.

\subsection{1989 Outburst vs 2015 Outburst}
\label{1989}
\citet{syunyaev91} report on observations of V404 Cyg (=GS \(2023+338\)) 1989 outburst by the Roentgen observatory.  They were able to perform a timing analysis with the HEXE data (\(15-200\) keV) when the source flux had decreased to a \(\sim \) 1 Crab level.  They found that the  power spectrum was dominated by a strong low-frequency noise between \(0.01-5\) Hz, approximated by a power-law  with a slope \(\sim  f^{-1} \) in the energy range \(20-100\) keV. Moreover, the same analysis performed on the \(20-35\) keV and \(65-180\) keV energy band showed similar power spectra in shape and amplitude. 

These results can be compared with our period 4 data, where the average flux level (\( \sim 2\) Crab in the \(20-300\) keV band) is close to the \citet{syunyaev91} observations. During this relatively low flux period, the SPI power spectrum is also dominated by low-frequency variability and its rms values remain unchanged with energy. However, the low-frequency slope is significantly steeper with a slope of \( -1.5\).

Power spectra analysis was also performed for the 1989 outburst of V404 Cyg by \citet{oosterbroek1997}, using \textit{Ginga} data in the \(1.2-36.8\) keV energy range. \citet{oosterbroek1997} made power spectra covering \(\sim 0.004-512\) Hz and \(\sim 0.004-8\) Hz based on the two available time-resolution data.  For both data sets, the model was comprised of two zero-centered Lorentzians to fit the low frequency regime and one non-zero-centered Lorentzian for the higher frequency portion of the power spectrum. The results consist of a large variety of best-fit parameters, even if the authors concluded that the shape of the power spectrum does not change significantly.  Our peak frequency and rms values are commensurable with the ranges observed in the \textit{Ginga} data.
On the other hand,  we note that many of the \textit{Ginga} power spectra present a flattening below \(\sim 0.1 \) Hz, we fail to detect in our data sets. If we ascribe that to the different energy ranges used in the analyses,
it could imply that different physical mechanisms are at work, and behave dissimilarly.

Considering the \textrm{rms} fractions, they range, in the \textit{Ginga} data sets, between 20 and 40\%  (\(0.01-8\) Hz) slightly higher than the values found in our analysis. Also, they appear stable  above 10 keV up to \(\sim\) 30 keV, in contrast with the significant decrease  observed above 50 keV in the SPI data set for periods 1 and 3.  Here too, it may be due to the energy domain or to the huge diversity and variability of the source characteristics.
 
\subsection{Comparison with Other Sources}  

\subsubsection{Power Spectra}
\label{othersources}
V404 Cyg reached uncommon luminosity levels and displayed an enormous flux variability at all wavelengths (e.g. \citet{kimura2016} and references therein).  The extreme levels of luminosity and variability were observed up to a few hundreds of keV (see Fig.~\ref{fig:lc}), which is rarely observed due to low numbers of photons at these energies. The BH transient V4641 Sgr showed a similar behavior during its 1999 outburst, though such an event lasted for only a few hours. Its peak flux was \(\sim 12\) Crab in the  X-ray band and the energy spectrum, observed up to \(\sim  200\) keV, suggested V4641 Sgr was in the hard state \citep{wijnands2000, revnivstev2002}.  Interestingly, power spectra analysis of the source in the \(2-22.1\) keV energy range shows no clear flat-top, though such a shape is generally expected for power spectra in the hard state.
The power spectrum is well described by a broken power-law with \( \Gamma_1 = -1.03\), a break at \(5.1\) Hz, and \( \Gamma_2 = -2.16\) \citep{wijnands2000} and  high rms  (up to 50\%), decreasing with energy (around 30\% between 20 and 60 keV).  These values are very similar to those found during our period 3, when we include a third power-law to account for an additional low-frequency component (see section \ref{bkn}).  The V404 Cyg power spectra for periods 1 and 2 are also well described by a broken power-law shape with comparable low frequency slopes (\(\Gamma_1 \sim -1.1\)) but flatter high frequency slopes (\(\Gamma_2 \sim -1.7\)). The break frequencies are consistent or lower by a factor of a few with that of V4641 Sgr (Table~\ref{table:l3} bottom panel), which could be important information on the source geometry. \\
Concerning the rms, the values obtained along the flare, between 15 and 30\%, are comparable to those obtained for Cyg X-1 in similar energy bands from  SPI data \citep{cabanac2011} or Suzaku data \citep{torii2011}. Moreover, the rms behavior with energy is changing in V404 Cyg.  For comparison, observations of Cyg X-1 by \citet{cabanac2011} found a decreasing rms with energy while \citet{torii2011} found a constant rms value from 20 to 200 keV.

\subsubsection{Time Lags}
Since \citet{miyamoto1988}, it has been known that the hard state X-ray spectrum cannot be due to a uniform Comptonizing medium, which predicts a roughly constant time lag.  Instead observations find a power-law dependence with longer time lags at lower frequencies.  The time lags range from milliseconds to a few seconds.  The slope of \( \Gamma = 0.79 \) seen in V404 Cyg is consistent with the \( \nu^{-0.8}\) that has been found at hard X-rays for Cyg X-1 \citep{crary1998} and GRO J\(0422+32\) \citep{grove1998}.  A study of many Cyg X-1 observations at lower energy \citep{pottschmidt2000} found also a similar trend, except during state transitions, implying that the lag behavior is independent of the source state. 

\subsection{Physical Interpretation}
\label{interpretation}
BLNs have been modeled as the result of shot-noise\footnotemark[1] processes resulting from such processes as magnetic flares or infalling blobs (see \citet{vanderklis2004} and references within). Other proposed models involved self-organized critical state processes in which the accretion disk evolves to and remains in an organized state independent of the initial conditions \citep{mineshige1994}. However, \citet{uttley2005} deduced, from the linear rms-flux relation observed in many sources, that fluxes should follow a log-normal distribution (which was supported by Cyg X-1 data) meaning that the variability process is multiplicative. Consequently, additive shot-noise models (where shots on all time-scales are independent) and self-organized criticality models can be ruled out as they cannot reproduce these features  \citep{uttley2005}.
\footnotetext[1]{Shot noise is a time-series model of random pulses (or shots), as suggested by \citet{terrellolsen1970}}

Another class of models, fluctuating-propagation models, have been developed (see for example \citet{lyubarskii1997}). They are multiplicative and thus are in agreement with a log-normal flux distribution.  In such models, instabilities in the accretion flow at large distances from the source result in fluctuations in the region near the source where the hard X-rays are emitted, with fluctuations likely due to magnetorotational instabilities \citep{hawley1991}.

textbf{In the framework of the fluctuating-propagation model, \citet{churazov2001} considered an optically thin corona extending to large radii above and below the optically thick disk. They propose that instabilities  occur at different (large) radii in the corona.  The instabilities propagate towards the BH and modulate the mass accretion rate through a given radius.  The product of the fluctuations from prior radii results in power-law shaped PDS over a wide range of frequencies \citep{lyubarskii1997}. Also, the flat-top features observed in the hard state are the result of damping by the truncated accretion disk. This damping occurs at the radius where the geometrically thin disk makes way for an optically thin accretion flow similar to the corona, typically below \( \sim 1\) Hz. 

The broken power-law shape seen in V404 Cyg periods \(1-3\) and V4641 Sgr can be interpreted within the same framework. However, the power spectra for V404 Cyg and V4641 Sgr lack the typical flat-top features despite both sources being in a hard state. For the 1989 outburst of V404 Cyg \citep{tanaka1995,zycki1999} and the 1999 outburst of V4641 Sgr \citep{revnivstev2002}, results show evidence for a dense outflow produced from ejected material and suggesting disruptions of the disk (see \citet{done2007} and references within). This unusual configuration could potentially arise from the sources being near or above the Eddington luminosity and reduce any damping effects from the disk. The extent to which the disk was disrupted would affect the frequency at which the flat-top begins.

With no flat-top detected in the SPI data down to \(10^{-2}\) Hz, a lower limit on the truncation radius can be estimated. Following the method employed in \citet{gilfanov2010} where the break frequency is equal to the inverse of the viscous time, a lower limit of the truncation radius is found to be \( \sim 2500 r_g\). Interestingly, \citet{zycki1999} estimate an unstable radius at \((5 - 8) \times 10^{10}\) cm (\( \sim 1200 - 2300 r_g\)), a value  commensurable with our above estimation.   

The fluctuating-propagation model can also explain the power-law dependence of the time lags.  In such models, the temporal variability also corresponds to small variability in the local spectra. Assuming the local spectra emit like a power law with the slope decreasing with decreasing distance from the BH \citep{kotov2001,arevalo2006}, the fluctuations at different distances in the accretion flow result in variability at different timescales being propagated towards the BH.  Moreover, recent works including Lense-Therring precession to account for QPOs have been able to model the power spectra of X-ray binaries using fluctuating-propagation models \citep{ingram2011,ingram2012,ingram2013}.

\section{Conclusion}
We analyzed the \(20-300\) keV power spectra for V404 Cyg during \textit{INTEGRAL} revolution 1557 (MJD \(57199.064-57200.991\)). The source intensity reached extreme levels up to a few hundred keV,  allowing timing studies in the hard X-ray domain.  Power spectra from three high SNR periods showed three BLN features similar to reported behavior from \textit{Fermi}/GBM observations during the same period \citep{jenke2016} and \textit{Ginga} observations from the 1989 outburst \citep{oosterbroek1997}.  Like those works, no QPOs were detected in the SPI data.\\

The high SNR periods could also be modelled by broken power-laws.  However, they do not show a clear flat-top typically seen in hard state power spectra.  Similar behavior was reported for the 1989 outburst of V404 Cyg as well as for a strong outburst in another BH transient, V4641 Sgr. The fluctuating-propagation model provides a global interpretation of the observed properties. In this framework, the atypical power spectra (and light curves) for both sources are potentially due to accretion disk disruption related to the sources emitting near the Eddington luminosity.  

A future perspective will be to explore the relation between spectral (disk) emission and timing properties as such an exceptional outburst conceals inestimable information about black hole physics and sub or super Eddington regimes.

\section*{Acknowledgments}  The \textit{INTEGRAL} SPI project has been completed under the responsibility and leadership of CNES.  We are grateful to ASI, CEA, CNES, DLR, ESA, INTA, NASA and OSTC for support.  JR acknowledges funding support CNES.

\end{document}